\documentclass[pra,amsfonts,twocolumn,nofootinbib,showpacs]{revtex4}
\RequirePackage{graphicx}
\usepackage{enumerate}
\usepackage{physics}
\usepackage{amsmath, amssymb}
\usepackage{slashed}
\usepackage{type1cm}
\usepackage{amsfonts}
\usepackage{bm} 
\usepackage[legacycolonsymbols]{mathtools} 
\begin{document}

\title{Exact WKB analysis for dynamical and geometric exponents in
generalized and nonlinear Landau-Zener transitions}

\author{Tomohiro Matsuda}
\affiliation{Laboratory of Physics, Saitama Institute of Technology,
Fukaya, Saitama 369-0293, Japan}

\begin{abstract}
The Berry phase is a geometric phase that is important in explaining
topological quantum phenomena.
The Berry phase is also important in non-perturbative phenomena, as the
imaginary part of the phase explains the non-perturbative transitions.
However, problems arose because the singular perturbation with respect
 to the Planck constant has not been treated adequately in conventional
 calculations, where the most serious problem is the arbitrariness of
 approximate calculations.
To solve this problem, we consider the exact WKB, which is a mathematical
 method that treats perturbative expansion with respect to the Planck
 constant as a rigorous singular perturbation.
This method is also a powerful computational tool
that makes analytical computation much easier for mathematical software.
Using the exact WKB, we analyze the derivation of the dynamical and the
 geometric exponents in generalized Landau-Zener models, highlighting
 the differences from other calculational methods. 
The discontinuity of complex geometric factor is a universal phenomenon
 that manifests itself in phase transitions, boundaries, particle
 generation, and topology changes.
These phenomena are ``non-perturbative'' in physics, while
mathematically, these discontinuities can be deeply related to the
singular structure of complex analysis.
The mathematical structure of these phenomena will be revealed
 by using the exact WKB.
\end{abstract}
\maketitle

\section{Introduction}
\hspace*{\parindent}
The general Wentzel-Kramers-Brillouin-Jeffreys (WKBJ or WKB) method is
known as an approximate solution to the Schr\"odinger equation and has
historically been used to understand quantum phenomena.
As a semiclassical approximation, the WKB method is convenient for
understanding the correspondence between quantization and the classical
treatment. 
The WKB method is also excellent for describing tunneling and other
typical quantum phenomena.
From a mathematical point of view, it was seen as an excellent way of
describing the mixing of eigenvalue equations or solutions of
differential equations (the Stokes phenomenon), which could be applied
to global problems.
On the other hand, the conventional WKB method is an approximation
useful for asymptotic expansion, and the solutions are not well defined in
the vicinity of the Stokes lines of the most interest\cite{WKB:review}.
For this reason, Voros\cite{Voros:1983} began trying to use the Borel
resummation to convert the WKB method into an exact analysis rather than
an approximation.
Technically, a technique called resurgence is used\cite{Voros:1983,
Delabaere:1993, Silverstone:2008, 
Pham:1988, CNP:1993, DDP:1993, DDP:1997, ExactWKB:textbook,
Virtual:2015HKT,RPN:2017} for the analysis.
The exact WKB considered in this paper is a rigorous analysis
based on the Borel resummation, developed by the group of authors of 
Ref.\cite{Voros:1983, Delabaere:1993,Silverstone:2008, Pham:1988, CNP:1993,
DDP:1993, DDP:1997, ExactWKB:textbook, Virtual:2015HKT}. 
More specifically, the Borel resummation is used to derive the connection
formula of the solutions at the Stokes lines extending from a turning
point.
Also, recent developments have established a deep and rigorous
connection between the exact WKB and topological
recursion\cite{Bouchard:2017}, which 
are two powerful frameworks used in mathematical physics and quantum
geometry. 
For physics, the most important aspect of discussing quantum theory
using the exact WKB is that the Stokes lines, which determine the global
properties of the solutions, are {\it exactly determined} with respect to the
expansion using the Planck constant.
In this sense, the ambiguity about the structure of the Stokes lines is
completely removed in the exact WKB.
As the structure of the Stokes lines reflects the structure of the theory
itself, it is crucial that there is no ambiguity here.
As approximations can easily alter the structure of these Stokes
lines\cite{Enomoto:2020xlf, Enomoto:2021hfv,Enomoto:2022nuj}, it is 
important to find approximations that do not change the basic structure of the
Stokes lines to preserve the qualitative structure of the theory.
This problem becomes particularly apparent when approximating the circle
using a quadratic polynomial\cite{Enomoto:2022nuj}.
To understand the specific issues involved, the second half of this
introduction will discuss the topic as it is presented in Berry's
original paper\cite{Berry:1990}.

The importance of geometric phases or exponent
factors was discussed by Berry\cite{Berry:1984jv,Berry:1990} and has
become an important research topic in condensed matter physics\cite{XCN:2010}.
As an adiabatic process, electrons acquiring a geometric phase are known
to provoke exotic effects such as quantum Hall effects\cite{TKNN:1982}.
Also, Berry's geometric phase is extended to the Uhlmann
phase\cite{Uhlmann:1986}, which is a generalization of the Berry phase
to mixed quantum states described by density matrices.
Compared to the study of geometric phases for adiabatic
processes, which has shown tremendous
progress\cite{Geometric:textbooks,Aharonov:1987gg, Wilczek:1984dh,
Samuel:1988zz, Mukunda:1991rc, Bohm:2003}, 
the study of geometric factors for  
non-adiabatic processes seems to be less advanced, except for some
cases\cite{Berry:1990, NR:1994, BKDSW:1996, TWO:2021, Chung:2024lky}.
Nevertheless, the non-perturbative discontinuity of the complex
Berry phase is a universal phenomenon that appears in all areas of
physics, such as phase transitions, particle creation, phase boundaries,
and topology changes.
As the calculational problem of the non-adiabatic process is often
linked to the problem of the WKB approximation, it is natural to expect that
the calculation with the exact WKB, which fixes the problem of the WKB
approximation, can make a significant contribution to this area.
The second half of this introduction will provide a more detailed
explanation of these issues.

In this paper, using the exact WKB, we derive the exact
geometric exponent factors of generalized Landau-Zener
models\cite{Landau:1932vnv, Zener:1932ws, Dykhne:1962, Davis:1976}. 
Our calculations of the exponents are correct for all orders and are
expected to be non-zero up to the second order of the geometric
exponent.\footnote{Berry's geometric exponent 
was derived using parallel
transport and does not depend on $\hbar$.}
We will compare our exact result with the previous result by
Berry\cite{Berry:1990}, which was derived up to the first order.
Prior to the exact calculation in Sec.\ref{sec-exact}, the ambiguity
of this first-order 
calculation will be manifested below.
We shall refer to our higher-order contribution as the ``quasi-geometric
exponent factor''.\footnote{The reason why the geometric exponent is
called ``geometric'' seems very clear in Berry's original
definition by parallel transport.
However, in our calculation using the exact WKB, it is just defined as
the coefficients of the $O(\hbar)$ expansion.}
We will also calculate the exponents exactly for all orders for the
original twisted Landau-Zener transition.

To confirm the connection between geometric meaning and
our calculation, it is useful to first remember Berry's
calculations of Ref.\cite{Berry:1990}.
The simplest manifestations of the geometric amplitude, which occur in
two-state system, is described by the Hamiltonian as
\begin{eqnarray}
\hat{H}(\tau)&=&
\begin{pmatrix}
Z(\tau)&X(\tau)-iY(\tau)\\
X(\tau)+iY(\tau)&-Z(\tau)\\
\end{pmatrix}\nonumber\\
&=&R(\tau)\begin{pmatrix}
\cos\theta(\tau)&\sin \theta(\tau) e^{-i\phi(\tau)}\\
\sin \theta(\tau) e^{+i\phi(\tau)}&-\cos\theta(\tau)\\
\end{pmatrix},
\end{eqnarray}
where polar (spherical) components $(R(\tau),\theta(\tau),\phi(\tau))$ 
are used for $(X,Y,Z)$.
This Hamiltonian has two eigenstates ($|u_\pm(\tau)\rangle$) with eigenvalues
\begin{eqnarray}
E_{\pm}(\tau)&=&\pm \sqrt{X^2(\tau)+Y^2(\tau)+Z^2(\tau)}.
\end{eqnarray}
The evolving state satisfying
\begin{eqnarray}
i\hbar \frac{d}{d\tau}|\psi\rangle&=&\hat{H}|\psi\rangle
\end{eqnarray}
will have an approximate solution
\begin{eqnarray}
|\psi(\tau)\rangle\simeq |u_+ \rangle \exp\left[-\frac{i}{\hbar}\int^\tau_0
E_+(\tau') d\tau'
\right].
\end{eqnarray}
Requiring parallel transport ($\langle u_+|\dot{u}_+\rangle=0$), 
the state evolution can be solved as
\begin{eqnarray}
|u_+(\tau)\rangle&=&
\begin{pmatrix}
\cos\left(\theta(\tau)/2\right))e^{-i\phi(\tau)/2}\\
\sin\left(\theta(\tau)/2\right))e^{+i\phi(\tau)/2}\\
\end{pmatrix}
e^{i\mu(\tau)},
\end{eqnarray}
where 
\begin{eqnarray}
\mu(\tau)&=&\frac{1}{2}\int_0^\tau d\tau' \dot{\phi}\cos\theta\\
&=&
\frac{1}{2}\int_0^\tau d\tau'
\frac{(X\dot{Y}-Y\dot{X})Z}{(X^2+Y^2)\sqrt{X^2+Y^2+Z^2}}.
\end{eqnarray}
As the exact WKB is not used here, the discontinuity at the Stokes lines
in the complex plane is not seriously considered.
Although not important for the present calculation, this calculation
also disregards a possible discontinuous phase change known as the Voros
factor\cite{Voros:1983}.
{\it The important point is that too many mathematical problems are ignored
in this calculation if it is looked at from a modern mathematical point of view.}

Then, the geometric exponent is calculated simply as the analytic
continuation of the phase $\mu(\tau)$, which gives
\begin{eqnarray}
\Gamma_g&=&-2\mathrm{Im}\int^{\tau_c}_0 d\tau \dot{\phi}\cos \theta,
\end{eqnarray}
 where $\tau_c$ is a simple zero of $X^2+Y^2+Z^2$.
The ambiguity surrounding the Planck constant may not be apparent here,
but it becomes clear when considering the following transformation.
\begin{eqnarray}
|\psi\rangle&=&\hat{U}(\tau)|\psi'\rangle\nonumber\\
&=&\begin{pmatrix}
e^{-i\phi/2}&0\\
0&e^{+i\phi/2}\\
\end{pmatrix}|\psi'\rangle,
\end{eqnarray}
where $|\psi'\rangle$ satisfies the equation in which Hamiltonian is
replaced by
\begin{eqnarray}
\hat{H}'&=&\begin{pmatrix}
Z-\frac{\hbar}{2} \dot{\phi}&\sqrt{X^2+Y^2}\\
\sqrt{X^2+Y^2}&-Z+\frac{\hbar }{2}\dot{\phi}&
\end{pmatrix}.
\end{eqnarray}
As will be described in detail in this paper, it is possible for the
exact WKB to determine exactly how the newly
appeared diagonal component ($\sim \hbar \dot{\phi}$) alters or keeps 
the global structure of the Stokes lines.
However, in conventional calculations, a non-trivial assumption must be made.
As this point is not often discussed, we will explain it in a little more
detail.
Here, what we are interested in is the consequence of 
$Z\rightarrow Z-\frac{\hbar}{2}\dot{\phi}$.
Put simply, the final results (i.e, the exponents) should not change as a
result of the above transformation. 
Therefore, one should consider the new terms proportional to $\hbar$ as
something that do not affect the previous calculation in the end.
Then, as is described by Berry in Ref.\cite{Berry:1990}, the original
geometric exponent ($\Gamma_g$) is calculated as a part (a higher term
$\propto \hbar^0$) of the new dynamical exponent $\Gamma_d'$.
The calculation was
\begin{eqnarray}
\label{eq-uniBe}
\exp(\Gamma_d')&=&\exp\left[
\frac{4}{\hbar}\mathrm{Im} \int^{\tau_c}_0 d\tau
E_+'(\tau)\right]\nonumber\\
&\simeq&\exp\left[
-\frac{4}{\hbar}\left|\mathrm{Im}\int_0^{\tau_c} d\tau R(\tau)\right|
-2\mathrm{Im}\int^{\tau_c}_0 d\tau \frac{\dot{\phi}Z}{R}
\right]\nonumber\\
&=&\exp\left[\Gamma_d + \Gamma_g\right]
\end{eqnarray}
Here, the higher term $\propto \hbar^1$ was neglected and the
consistency of this term was not discussed.
The important point to note here is that the derivation of the new
zeros, which should be $\tau_c'\ne \tau_c$ 
after $Z\rightarrow Z-\frac{\hbar}{2}\dot{\phi}$,
has {\it not} been taken into account.
In short,
$\tau_c'$ for the new $E'$ was just replaced by $\tau_c$.
In fact, if $\tau_c$ changes in the new Hamiltonian, the results will
be changed and it makes the above consistency of the transformation 
quite unclear.
We will show explicitly that such problem arises in the spiral-type
Landau-Zener model using cylinder coordinates.
This suggests the presence of a computational flaw in the conventional
calculation. 
However, this does not mean that there is anything wrong with the
concept of geometric factors.
What is important here is that the consistency of the calculation
can only be explained with the exact WKB, for which the expansion is exact.
We are afraid that it has not been widely understood that ``conventional
calculations'' are based on many empirical rules and have no rigid
mathematical basis especially for the $\hbar$ expansion.
{\it Our claim is that a rigorous mathematical framework is inevitable
 when giving physical meaning to these calculations.
The aim of this paper is to organize this part of the discussion in
physics.}
We will often return to this point later with more specific examples.

We will now provide an intuitive explanation of the twisted Landau-Zener
transition.
Using the vector spin-$\frac{1}{2}$ operator $\hat{\bf S}$ and Hamiltonian
vector ${\bf H}=(X,Y,Z)$, we have $\hat{H}(\tau)={\bf H\cdot \hat{S}}$.
If the Hamiltonian vector is given by
\begin{eqnarray}
{\bf H}(\tau)&=&(\Delta \cos\phi(\tau), \Delta \sin\phi(\tau),A\tau),
\end{eqnarray}
the Hamiltonian curve lies on a cylinder centered on the $Z$ axis.
In this model, the conventional Landau-Zener model corresponds to
$\phi=0$ (i.e, without rotational motion on the cylinder).
As might be imagined immediately, a simple spiral motion is produced if
$\phi\propto \tau$, while a ``twist'' that changes direction appears 
if $\phi\propto t^2$.

The original twisted Landau-Zener model is twisted once by a
quadratic term.
In this case, in the geometric exponent, there is no correlation 
between the positions of the twist and the Landau-Zener transition.
Since the geometric factor is proportional to the second
derivative of the angular function ($\ddot{\phi}$) in this model, the
geometric factor becomes independent of location of the twist
because $\ddot{\phi}$ is a constant. 

In this paper, to demonstrate that 
the exact WKB is not only exact but is also very 
convenient for the calculation of complex models, we
consider a double-twisted model with a cubic term in Sec.\ref{sec-DT} 
and compute the exponents for all orders.
Such calculation must be impossible without the exact WKB.

Furthermore, in Sec.\ref{sec-NL}, we show that the exact WKB can also be used to
calculate the exponents of the Landau-Zener transition with non-linear 
diagonal elements.
Specifically, we consider a model with the diagonal
elements proportional to $t^2$.
Typically, calculations are performed by considering linear expansions
centered on the intersection points.
{\it Using the exact WKB, we point out the critical flaw in this calculation.}
To demonstrate the importance of the calculation, we examine the
validity of the conventional linear approximation at the intersection,
comparing it with the exact calculation.
{\it Our calculations show that the naive linear approximation is quite
unreliable.}
The difference between the conventional linear approximation and the
exact calculation is illustrated in Fig.\ref{fig-kappa2}.
This difference is very important in physics because, in actual
experiments, level-crossings will be accompanied with non-linear terms.
For the non-linear Landau-Zener model, using the Stokes lines of the
exact WKB, we also show that the ``phantom'' Landau-Zener transition occurs
without the intersection of the diagonal elements. 
We show that the exact WKB is not only rigorous but also a very useful
calculational tool for these models.
It plays a crucial role in revealing the mathematical structure
behind the phenomena.

The geometric effects in the Landau-Zener transition are known to lead to
nontrivial dynamics of electrons in Dirac and Weyl semimetals driven by
strong electric laser fields.
Using the Pauli matrix $\hat{\sigma}^j, (j=x,y,z)$,
the effective Hamiltonian for the fermion with chilarity $\xi=\pm$ is
given by\cite{TWO:2021}
\begin{eqnarray}
\hat{H}&=&v \left[
\xi(k_x+e A_x)\hat{\sigma}^x
+(k_y+e A_y)\hat{\sigma}^y
\right]+m\hat{\sigma}^z,
\end{eqnarray}
where $e,v,m$ are the electric charge, the Fermi velocity and the mass
parameter, respectively.
This model has implications to valleytronics in 2D
materials\cite{YXN:2008, XLFXY:2012}.
The model can be extended to 3D massless Dirac fermions subject to
rotating electric fields\cite{TWO:2021}.
The twisted Landau-Zener model can be mapped into the twisted Schwinger
model, which is realized in rotational electric fields and the
electron-hole pairs with the Hamiltonian\cite{TWO:2021}.
Let us take a closer look at the quantization of the electromagnetic
field in this model.
Details will be described in Sec.\ref{sec-Q}.
Let us start with the standard optical absorption process, where a perturbative
picture is employed\cite{TWO:2021,YXN:2008,XLFXY:2012}.
Then, electrons in the occupied bands are excited to the unoccupied
bands with the energy difference given by the photon energy.
On the other hand, in the case of tunneling excitations, the creation of
the electron and hole pairs is not driven by the perturbative photon
absorption process but by the nonadiabatic
process\cite{TWO:2021,YXN:2008,XLFXY:2012}.
Here, the twisted Landau-Zener transition explains the non-adiabatic (and
non-perturbative) tunneling excitations.
Our question is straightforward. 
Why does such a difference arise when these two processes should be
described by the same equation? 
We approach this problem from the perspective that ``quantization'' may
alter the properties of the equation.
It is highly peculiar that the same equation could exhibit different
properties. 
However, we note that the exact WKB achieves precisely this by
introducing explicit $\hbar$-dependence to a parameter.
We will introduce an explicit Planck constant in the model, expecting a
unified calculation of perturbative photon absorption and the
non-adiabatic tunneling.
We show that the dynamical exponent is changed by
the ``quantization'' (introduction of an explicit $\hbar$) 
as the Stokes lines are changed if it is analyzed by the exact
WKB.\footnote{Ref.\cite{NR:1994} takes into account 
this contribution using a unitary transformation. 
With the exact WKB, these calculations are unified.} 
This means that the phenomenon of the Landau-Zener transition being
enhanced by the absorption of photons can be seen as a natural consequence
of the quantization.
Note however that the term ``quantization'' is used here in the very 
restricted sense that the
underlying physical quantity is replaced by a quantity that includes
the Planck constant explicitly.
To be more specific, we assume $\epsilon=\hbar \omega$, and $\epsilon$ is
fundamental after quantization while $\omega$ is fundamental for the
classical electromagnetic field.
Although the fact that the properties of equations can be altered by such
differences in the definition of fundamental parameters has not been
widely discussed in the past, this is essential for the quantum
dynamics.
The calculations thus far have been far too sloppy on this point.

As described above, the exact WKB used in this paper is not merely a
mathematical 
calculation technique, but the only tool that has the potential to 
approach the essence of physics that has been dismissed in other
calculation methods.

\section{Introduction to the exact WKB}
\begin{figure}[ht]
\centering
\includegraphics[width=0.8\columnwidth]{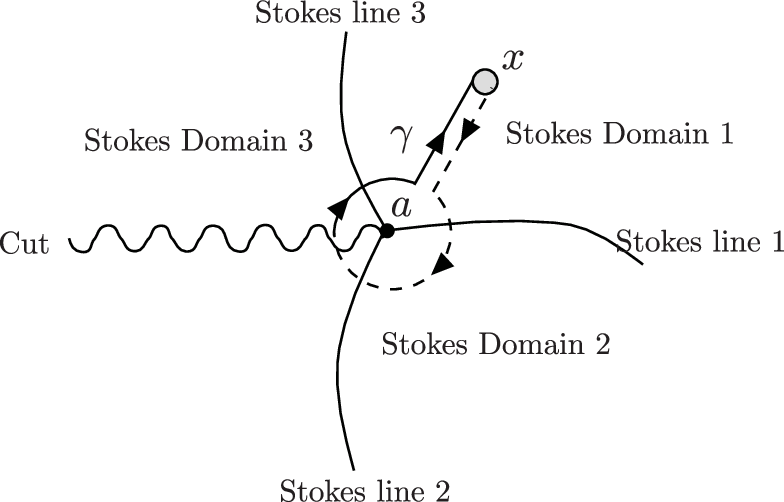}
 \caption{The diagram of the contour integral $\gamma$ around the
 turning point $a$. Stokes lines and the contour are drawn on the
 complex $x$-plane.}
\label{fig_contour}
\end{figure}

While the explanation in this section largely overlaps with that in
previous papers\cite{Enomoto:2020xlf, Enomoto:2021hfv} and
textbooks\cite{ExactWKB:textbook, Virtual:2015HKT, RPN:2017}, we will
provide a somewhat redundant 
explanation here for the convenience of readers who are not familiar
with exact WKB and the Stokes phenomenon.
Although the exact WKB is now capable of handling higher order ordinary
differential equations\cite{Virtual:2015HKT,Enomoto:2022nuj}, all
analyses are attributed to the ``Schr\"odinger equation'' given by
\begin{eqnarray}
\left(-\frac{\hbar^2}{2}\frac{d^2}{dx^2}+U(x)\right)\psi(x)&=&E\psi(x).
\end{eqnarray}
Here, following the Schr\"odinger equation in physics, $U(x)$ is called
``potential'' and $E$ is called ``energy''.
Note that in mathematics these definitions may vary slightly depending on the situation.
This $\hbar$ is the Planck constant and the conventional WKB method
considers the expansion by small $\hbar$.
However, as can be seen from the fact that 
this equation is no longer a differential equation at $\hbar=0$,
 the WKB method is not just an approximation but what is mathematically
 known as ``singular perturbation''.
If the singular perturbation is taken into account, $\hbar$ is
necessarily continued analytically in the complex plane.
For the exact WKB, it is common to consider $\eta=\hbar^{-1}$ instead of
the Planck constant, as it is convenient for the Borel transform.
For this reason, $\eta$ is used in this paper, according to
mathematical notation.
This introduction also uses the following simpler equation as the basic
formula,
\begin{eqnarray}
\left(-\frac{d^2}{dx^2}+\eta^2 Q(x,\eta)\right)\psi(x)&=&0.
\end{eqnarray}
$Q(x,\eta)$ in the above equation is
sometimes referred to as ``potential''.
Notice that there are two variables, $x$ and $\eta$.
The solutions of the above equation will be considered on the complex
plane of both $x$ and $\eta$.
In general, $Q(x,\eta)$ can be regarded as a polynomial, but it is known
that the connection formula at a Stokes line is similar for rational
functions, as long as the behavior of singular points is
properly considered.
The solution to the above equation is given in the following form,
\begin{eqnarray}
\psi(x,\eta)&=&\exp\left(\int^x S(x',\eta) dx'\right).
\end{eqnarray}
Note here again that $\eta$ is considered as a variable.
The expansion is a series with respect to this $\eta$ and is considered
as follows.
\begin{eqnarray}
S(x,\eta)&=&\eta S_{-1}(x)+S_0(x)+\eta^{-1}S_1(x)+\cdots
\end{eqnarray}
The first step is to construct the WKB solution in the same way as the
standard WKB approximation.
The equation satisfied by $S(x,\eta)$ is the following Riccati equation,
\begin{eqnarray}
S^2+\frac{dS}{dx}&=&\eta^2 Q.
\end{eqnarray}
Strictly speaking, the $x$ derivative above is a partial derivative with
respect to $S(x,\eta)$, but we will follow the standard
convention.
For our later calculations, consider the following expansion of
$Q(x,\eta)$;
\begin{eqnarray}
Q(x,\eta)&=&Q_0(x)+\eta^{-1}Q_1(x)+\eta^{-2}Q_2(x).
\end{eqnarray}
This is all that is needed to describe the solutions in this paper, as
higher terms will not 
appear in $Q(x,\eta)$ of the Landau-Zener model.
The following equations are obtained by comparing each order,
\begin{eqnarray}
S_{-1}^2&=&Q_0,\\
2S_{-1}S_0+\frac{dS_{-1}}{dx}&=&Q_1,\\
2S_{-1}S_1+S_0^2+\frac{dS_{0}}{dx}&=&Q_2,\\
2S_{-1}S_n+\sum_{j=0}^{n-1}S_jS_{n-j}+\frac{dS_{n-1}}{dx}&=&0
\,\,\,\,(n\ge 2).
\end{eqnarray}
Two different solutions with $(+)$ and $(-)$ signs can be constructed by
putting the first solution 
as $S_{-1}^{(\pm)}=\pm\sqrt{Q_0}$.
If we introduce
\begin{eqnarray}
S_{odd}&\equiv&\frac{S^{(+)}-S^{-}}{2}\\
S_{even}&\equiv&\frac{S^{(+)}+S^{-}}{2},
\end{eqnarray}
the above Riccati equation gives
\begin{eqnarray}
S_{even}(x,\eta)&=&-\frac{1}{2}\log S_{odd}(x,\eta).
\end{eqnarray}
Therefore, we have
\begin{eqnarray}
\int^x S^{(\pm)}(x',\eta)dx'&=&\pm \int^x
 S_{odd}(x',\eta)dx'\nonumber\\
&&-\frac{1}{2}\log S_{odd}(x,\eta),
\end{eqnarray}
which leads to the WKB solution
\begin{eqnarray}
\psi_{\pm}(x,\eta)&=&\frac{1}{\sqrt{S_{odd}(x,\eta)}}\exp\left(\pm
\int^x S_{odd}(x',\eta)dx'
\right).\nonumber\\
\end{eqnarray}
We shall obtain a specific form of $S_{odd}$, as it will be used later
in our calculation of geometric exponents.
We have
\begin{eqnarray}
S_{-1}^{(\pm)}&=&\pm \sqrt{Q_0}\\
S_{0}^{(\pm)}&=&\frac{Q_1-d S_{-1}^{(\pm)}/dx}{2S_{-1}^{(\pm)}}\\
S_{1}^{(\pm)}&=&\frac{Q_2-d S_0^{(\pm)}/dx-\left(S_0^{\pm}\right)^2}{2S_{-1}^{(\pm)}},
\end{eqnarray}
which lead to
\begin{eqnarray}
\label{eq-sodd-expand}
S_{odd,-1}&=&\sqrt{Q_0}\\
S_{odd,0}&=&\frac{Q_1}{2\sqrt{Q_0}}\\
S_{odd,1}&=&\frac{Q_2}{2\sqrt{Q_0}}-\frac{5}{32}\frac{Q_0^{'2}}{Q_0^{5/2}}
+\frac{Q_0^{''}-Q_1^2}{8Q_0^{3/2}}.
\end{eqnarray}
Here, for a simple turning point $x=a$, the solution normalized by $a$
is defined in the following way,
\begin{eqnarray}
\psi_{\pm}(x,\eta)&=&\frac{1}{\sqrt{S_{odd}(x,\eta)}}\exp\left(\pm
\int^x_a S_{odd}(x',\eta)dx'
\right).\nonumber\\
\end{eqnarray}
Furthermore, the integral is redefined as a contour integral as
shown in Fig.\ref{fig_contour}, taking into account the cut around the
turning point,
\begin{eqnarray}
\int^x_a S_{odd}(x',\eta)dx'&=&\frac{1}{2}\int_\gamma S_{odd}(x,\eta)dx.
\end{eqnarray}
Having defined the integral of the solution by the contour integral, the
integral connecting the turning points that appears in the following
calculations will also be defined by the contour integral.
The following is a more detailed description of the treatment of
the exact WKB solutions, which can be skipped if the 
reader is only interested in the exact WKB as a useful calculational
method. 

The exact WKB uses the Borel resummation of the solutions in the
following way.
The Borel resummation can be thought of simply as ``a method of
describing divergent power series in terms of an integral''.
Specifically, the following transform is considered as the Borel
transform ($\simeq$ inverse Laplace transform).
\begin{eqnarray}
\label{Eq-Borel-trans}
&&f(\eta)=e^{s\eta}\sum_{n=0}^{\infty}\eta^{-n-\alpha}f_n\\
&\rightarrow&f_B(y)=\sum_{n=0}^{\infty}\frac{f_n}{\Gamma(n+\alpha)}(y+s)^{n+\alpha-1},
\end{eqnarray}
where $s,f_n\in \mathbb{C},\alpha\in
\mathbb{C}\backslash\{0,-1,-2,\cdots\}$.
The $\Gamma$ function in the denominator is important for the
 convergence of $f_B$.
The original function is reproduced by the Borel integral ($\simeq$
Laplace integral).
In summary, the flow of transform in the Borel resummation is
as follows.
\begin{eqnarray}
f(\eta)&\rightarrow& f_B(y)\rightarrow F(\eta)=\int^\infty_{-s}e^{\eta
 y}f_B(y)dy.
\end{eqnarray}
This replaces $f(\eta)$, which is a divergent power series, by
$F(\eta)$, with the integral of the convergent series $f_B$.  
In the above Eq.(\ref{Eq-Borel-trans}), the factor
$e^{s\eta}$ will play a very important role in the exact WKB.
This factor implies the first term of the exponent ($s(x)=\int^x
S_{odd,-1}dx'$) determines the endpoint of the $y$ integral.
The inverse Laplace transform may replace a divergent
power series by an ordinary series.
Then the function is called ``Borel summable''.
(Here we have used the term very loosely.
This is not the strict mathematical definition.) 
Once the divergent WKB solution is replaced by an ordinary series by
the Borel transform, and then restored using the Borel integral,
the function that was originally a divergent power series can be
replaced by a useful integral function. 
The poles that appear in the $y$-plane will require attention, and
this is exactly what is needed to understand the Stokes phenomenon.
Note that moving $\psi(x,\eta)$ on the complex $x$-plane causes $s(x)$
to move on the complex $y$-plane.
In this section, we will not go into the details of the derivation of the
connection formula.
Instead, we will explain the practical calculation of the exact WKB
using an example of the typical MTP structure (Merged pair of Turning
Points).
A more detailed explanation as application to particle production
in cosmology can be found in Ref.\cite{Enomoto:2020xlf,Enomoto:2021hfv},
which should be useful for understanding the calculation in this paper.

\begin{figure}[ht]
\centering
\includegraphics[width=1.0\columnwidth]{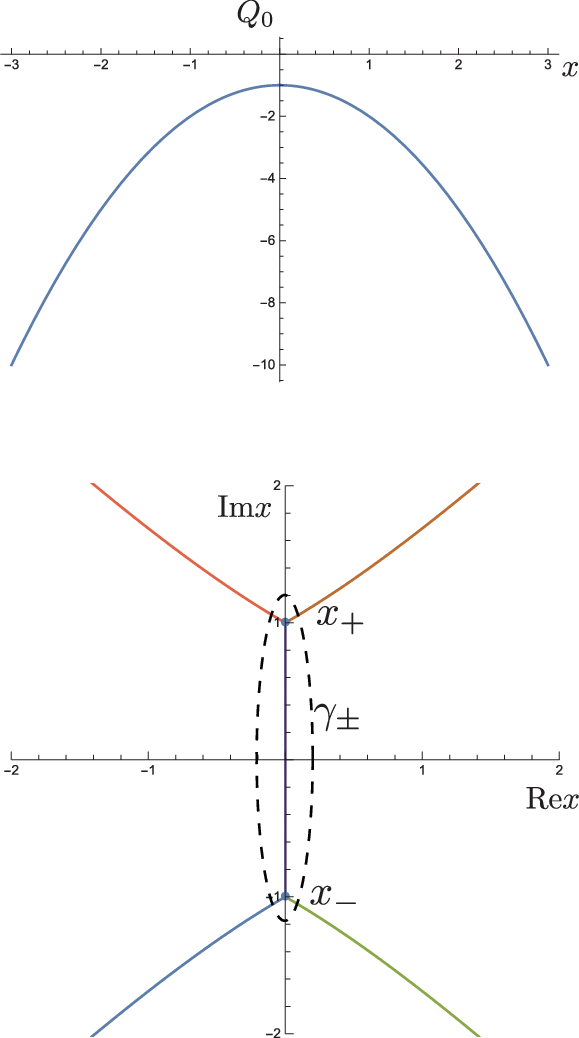}
 \caption{$Q_0(x)=1-x^2$ as a ``scattering potential'' is shown in the
 upper panel on the $Q-x$ plane, where the horizontal axis ($x$) is real.
The Stokes lines are written on the complex $x$-plane in the lower panel.
Colors are just dicriminating the three Stokes lines extending from a
turning point. 
These Stokes lines are called the MTP structure because the Stokes lines
 extending from the two turning points are degenerated and a pair of the
 turning points is merged.
The dynamical exponent appears when the solution crosses the Stokes
 line at the origin on the real axis. 
The exponent is calculated by the integration in Eq (\ref{Eq-MTP-dynexp}).
The contour of the integration is shown in the lower panel.}
\label{fig-MTP}
\end{figure}

First, let us consider the simplest model of the MTP structure.
When $Q_0=-E-\alpha x^2$, the turning points are the two solutions
$x_\pm=\pm i\sqrt{E/\alpha}$ of $Q_0(x)=0$.
The integral 
$\int_{x-}^{x+} S_{odd}dx=\frac{1}{2}\int_{\gamma\pm}S_{odd}dx$
 is a contour integral around two turning points, as is shown in Fig.\ref{fig-MTP}.
Here the dynamical exponent calculated by Berry in
Ref.\cite{Berry:1990} is
\begin{eqnarray}
\label{Eq-MTP-dynexp}
\Gamma_d&\equiv&-\Re \int_{\gamma\pm}S_{-1}dx\\
&=&-\Re \left[\pi
	 i\eta\underset{x=\infty}{\Res}\left[S_{-1}\right]\right]
=\frac{\pi E}{2\sqrt{\alpha}}\eta. 
\end{eqnarray}
From now on, $S_{odd,n}$ will be referred to as $S_{n}$ unless there is
a risk of confusion. 
To elaborate on the calculations here, the series expansion at $x\rightarrow \infty$ is
\begin{eqnarray}
\sqrt{-A^2 x^2-C^2}&\underset{x\rightarrow \infty}{\simeq}& 
i A x +\frac{i C^2}{2 A x}-\frac{i C^4}{8 A^3 x^3}\nonumber\\
&&+O\left(\left(\frac{1}{x}\right)^5\right),
\end{eqnarray}
which gives
\begin{eqnarray}
\underset{x=\infty}{\Res}\left[\sqrt{-A^2 x^2-C^2}\right]
&=&-i\frac{C^2}{2A}.
\end{eqnarray}
The above simplification of the calculation by the residue is one of the
features of this basic model.
 As the calculation using the residue is just a calculation
procedure, the same result can be obtained by considering the standard
local integral.
In the calculation with the exact WKB, the Stokes lines are exactly
determined only by $Q_0[x]$.
This is the mathematical consequence of the singular perturbation.
On the other hand, the exponent is
\begin{eqnarray}
\Gamma_{total}&=-\Re&\int_{\gamma\pm}S_{odd}dx,
\end{eqnarray}
which shows that contributions of the higher-order terms will appear in
the exponents.
Note that Berry's original definition\cite{Berry:1990} used the
imaginary part of the integral, whereas in the exact WKB, it is more
natural to define it as the real part of the above integral. 
These are the same thing (except for the difference in their derivation).
This simple model does not produce a geometric exponent factor.
In the calculation that follows, the geometric exponent is obtained from
the integral of $S_0$.
We will see that the result strongly depends on the functional form of $Q_1$.
To write it down specifically, the ``geometric exponent'' defined by
Berry is given by
\begin{eqnarray}
\Gamma_g&\equiv& -\Re \int_{\gamma\pm}S_{0}dx\\
&=&-\Re\left[\pi
 i\underset{x=\infty}{\Res}\left[S_{0}\right]\right]\nonumber\\
&=&-\Re \left[\pi
 i\underset{x=\infty}{\Res}\left[\frac{Q_1}{2\sqrt{Q_0}}\right]\right],
\end{eqnarray}
which can be non-zero only when $Q_1\ne 0$.
Also, it vanishes when the function is regular at complex infinity.
For a generalized Landau-Zener transition, in which the off-diagonal
elements are not constants, one may have a higher contribution (we call
it a ``quasi-geometric exponent'' in this paper),
\begin{eqnarray}
\Gamma_{g2}&\equiv&-\Re\int_{\gamma\pm}\eta^{-1}S_{1}dx\\
&=&-\Re\left[
\pi i\eta^{-1}\underset{x=\infty}{\Res}\left[S_{1}\right]\right].
\end{eqnarray}
It can be proven inductively that no higher terms can emerge in
generalized Landau-Zener (two-level) models.\footnote{What
one has to do for the proof is 
just to find that such higher terms are regular at $x=\infty$.}
Therefore, for the Landau-Zener model, these are all the exponential
factors to be calculated.
In this paper, these three ``dynamical'', ``geometric'' and
``quasi-geometric'' exponents are to be calculated in the Landau-Zener
and generalized Landau-Zener models.

\section{The exact WKB for exponents in the Landau-Zener transition}
\label{sec-exact}
We first review the original Landau-Zener model and
explain how to derive Berry's exponents using the exact WKB.
For completeness, the solutions with special functions are first described in
rather detail.

We start with a pair of first-order differential equations below
\begin{eqnarray}
i\hbar\frac{d}{dt}\left(
\begin{array}{c}
\psi_1\\
\psi_2
\end{array}
\right)&=&\left(
\begin{array}{cc}
-\frac{a}{2}t& \Delta \\
 \Delta& +\frac{a}{2}t 
\end{array}
\right)
\left(
\begin{array}{c}
\psi_1\\
\psi_2
\end{array}
\right),
\end{eqnarray}
where $a>0$ is a real parameter.
As we are considering the original Landau-Zener model, we assume here
that $\Delta$ is a real constant. 
Decoupling the equations, we have
\begin{eqnarray}
\left[\frac{d^2}{dt^2}+\frac{1}{\hbar^2}\left(\Delta^2+\frac{1}{4}a^2t^2\right)-\frac{i}{\hbar}\frac{v}{2}\right]\psi_1&=&0 \\
\left[\frac{d^2}{dt^2}+\frac{1}{\hbar^2}\left(\Delta^2+\frac{1}{4}a^2t^2\right)+\frac{i}{\hbar}\frac{v}{2}\right]\psi_2&=&0.
\end{eqnarray}
For the exact WKB, these are
\begin{eqnarray}
\label{eq-weberstokes}
Q_0&=&-\Delta^2-\frac{1}{4}a^2 t^2\nonumber\\
Q_1&=&\pm i\frac{a}{2}\nonumber\\
Q_2&=&0.
\end{eqnarray}
Let us first calculate the connection matrix using the exact solutions
based on the Weber functions.
Defining $z=i\sqrt{a} e^{i\pi/4}t$ ($z^2=-ivt^2$), we have
\begin{eqnarray}
\left[\frac{d^2}{dz^2}+\left(n+\frac{1}{2}-\frac{1}{4}z^2\right)\right]\psi_1(z)&=&0.
\end{eqnarray}
Here we defined
\begin{eqnarray}
n&\equiv&i\frac{\Delta^2}{a}.
\end{eqnarray}
This is nothing but the standard equation of the Weber function.
The solutions are given by a pair of independent functions among 
$D_n(z), D_n(-z), D_{-n-1}(iz), D_{-n-1}(-iz)$.
Using mathematics, one can easily find the
connection matrix connecting the left-hand states ($\psi^L$) to the
right-hand states ($\psi^R$) as\footnote{More details can be found in
Ref.\cite{Enomoto:2020xlf,Enomoto:2021hfv}}
\begin{eqnarray}
\left(
\begin{array}{c}
\psi^R_1\\
\psi^R_2
\end{array}
\right)&=&\left(
\begin{array}{cc}
e^{-\pi \kappa}e^{i\theta_1}& -\sqrt{1-e^{-2\pi\kappa}}e^{i\theta_2} \\
\sqrt{1-e^{-2\pi\kappa}}e^{-i\theta_2} & e^{-\pi \kappa}e^{-i\theta_1}
\end{array}
\right)
\left(
\begin{array}{c}
\psi_1^L\\
\psi_2^L
\end{array}
\right),\nonumber\\
\end{eqnarray}
where we introduced $\kappa\equiv\frac{\Delta^2}{a}$.
Let us consider the above calculation with the exact WKB.
Using the Stokes lines defined by Eq.(\ref{eq-weberstokes}), one will find
\begin{eqnarray}
-\pi \kappa&=&\int_{-\frac{2\Delta}{a}}^{+\frac{2\Delta}{a}} S_{odd}dt.
\end{eqnarray}
What is surprising about the exact WKB is that this simple calculation
 is shown to be exact even if the Landau-Zener transition is 
extended to include non-trivial $Q_1(t)$ and $Q_2(t)$.
To understand why this is the case, one has to
understand how the starting point of the Borel integral was determined
 by $S_{-1}$ {\it alone} when defining the Borel resummation, and how the
connection formula was subsequently derived using it.
Then, to understand what the ``surprise'' means, one has to try 
formulating the solution using special functions when $Q_1(t)$ and
 $Q_2(t)$ are non-trivial functions.

Here Berry's dynamical exponent is calculated as
\begin{eqnarray}
\Gamma_d&\equiv&- 2 \pi \kappa,
\end{eqnarray}
since $\Gamma_d$ is defined by the probability
 $P\simeq \exp[-\Gamma_d]$, not directly by the Bogoliubov coefficients.
We also have the phase factors $\theta_1$ and $\theta_2$, but they will
be neglected for our calculation.\footnote{For the exact WKB, the
phases are derived using the ``Voros factor'', not solely from the
conventional integral $\int^x S_{odd} dx'$.
The Voros factor\cite{Voros:1983} was originally introduced to resolve the apparent
discontinuity between $\eta=\eta_0 e^{\pm i\epsilon}$\cite{Voros:1983}.
See also the recent developments in Ref.\cite{Voros-ASU:2022,Voros-AU:2022}.}  
Note however that $\psi_{1,2}$ are not the adiabatic states.
Therefore, one can see that the ``transition'' for these functions
 becomes significant when $v\rightarrow 0$.
To avoid confusion, the above matrix has to be given for the adiabatic
states, which are not identical to
$\psi_{1,2}$.\footnote{Note that usually the ``adiabatic
energy'' $E_\pm=\pm\sqrt{\Delta^2+a^2t^2/4}$ is used for these states.}
We thus define the adiabatic states to have the matrix given by
\begin{eqnarray}
\label{eq-transmat-basic}
\left(
\begin{array}{c}
\Psi_1^+\\
\Psi_2^+
\end{array}
\right)&=&\left(
\begin{array}{cc}
 \sqrt{1-e^{-2\pi\kappa}} &e^{-\pi \kappa}\\
e^{-\pi \kappa} &-\sqrt{1-e^{-2\pi\kappa}}
\end{array}
\right)
\left(
\begin{array}{c}
\Psi_1^-\\
\Psi_2^-
\end{array}
\right),\nonumber\\
\end{eqnarray}
where phase parameters are neglected.
As can be seen immediately, the geometric exponent is zero in this
original Landau-Zener transition because $Q_1$ is a constant.

The above first calculation using special functions may appear very simple.
However, if $Q(x,\eta)$ has higher terms 
($Q_1(x)$ and $Q_2(x)$), and if they are non-trivial functions of $x$,
transforming the equation into the standard form for which special
functions can be used is generally a very tedious task.
On the other hand, with the exact WKB, the algebraic calculations
mentioned above can be carried out very easily for the integral of $S_{odd}$.
Since the Stokes lines are determined by $Q_0$, the
transition matrix is given in the same form (\ref{eq-transmat-basic})
even if $Q_1(x)$ and $Q_2(x)$ are non-trivial.
The higher-order terms are known to contribute only to the integration
in $\kappa$ as $\int_{\gamma\pm}S_{odd}dx$, and also, the integral is
simply given by the residue of the complex infinity as far as the
diagonal element is a linear function.
This makes it easier to prove inductively that contributions from
higher-order terms ($S_{odd,n},n\ge 2$) disappear from $\int_{\gamma\pm}
S_{odd}dx$, as what one has to do for the proof is 
just to find that such higher terms are regular at complex
$x$-infinity.

\subsection{The exact WKB for generalized Landau-Zener models}
Next, we will try to use the exact WKB on a model with an extension of
the original Landau-Zener transition. 
For the case where the non-diagonal element $\Delta(t)$ is explicitly
time-dependent, the transformation from the Landau-Zener transition to the
``Schr\"odinger equation'' must be derived at the first place. 
We introduce $D(t)$ and $\Delta(t)$ and assume 
\begin{eqnarray}
\label{eq-simpleoriginalLZ}
i\hbar \frac{d}{dt}\left(
\begin{array}{c}
X\\
Y
\end{array}
\right)&=&\left(
\begin{array}{cc}
D(t) & \Delta(t)^*\\
\Delta(t) & -D(t)
\end{array}
\right)
\left(
\begin{array}{c}
X\\
Y
\end{array}
\right).
\end{eqnarray}
Decoupling these equations, one will have
\begin{eqnarray}
\ddot{X}-\frac{\dot{\Delta}^*}{\Delta^*}\dot{X}+
\left(-\frac{iD\dot{\Delta}^*}{\hbar\Delta^*}
+\frac{i\dot{D}}{\hbar}
+\frac{|\Delta|^2+D^2}{\hbar^2}
\right)X=0.\\
\ddot{Y}-\frac{\dot{\Delta}}{\Delta}\dot{Y}+
\left(\frac{iD\dot{\Delta}}{\hbar\Delta}
-\frac{i\dot{D}}{\hbar}
+\frac{|\Delta|^2+D^2}{\hbar^2}
\right)Y=0.
\end{eqnarray}
To obtain the standard ``Schr\"odinger equation'' of the exact WKB, we
have to remove the second term that is proportional to $\dot{X}$ (or $\dot{Y}$).
Let us introduce new $\hat{X}$ and $\hat{Y}$ defined by
\begin{eqnarray}
\label{eq-normalEWKBtrans}
\hat{X}&=&\exp\left(-\frac{1}{2}\int^t
	       \frac{\dot{\Delta}^*}{\Delta^*}dt\right)X\nonumber\\
\hat{Y}&=&\exp\left(-\frac{1}{2}\int^t
	       \frac{\dot{\Delta}}{\Delta}dt\right)Y,
\end{eqnarray}
which recovers the standard formula of the exact WKB,
\begin{eqnarray}
\label{Eq_2ndorder-Majorana}
\ddot{\hat{X}}&+&\left(\frac{-iD\dot{\Delta}^*}{\hbar \Delta^*}
+\frac{i\dot{D}}{\hbar}+\frac{|\Delta|^2+D^2}{\hbar^2}\right.\nonumber\\
&+&\left.\frac{\ddot{\Delta}^*}{2\Delta^*}-\frac{3(\dot{\Delta}^*)^2}{4(\Delta^*)^2}\right)\hat{X}=0\\
\ddot{\hat{Y}}&+&\left(\frac{iD\dot{\Delta}}{\hbar \Delta}
-\frac{i\dot{D}}{\hbar}+\frac{|\Delta|^2+D^2}{\hbar^2}\right.\nonumber\\
&+&\left.\frac{\ddot{\Delta}}{2\Delta}-\frac{3(\dot{\Delta})^2}{4(\Delta)^2}\right)\hat{Y}=0.
\end{eqnarray}
Therefore, for $X$ we have 
\begin{eqnarray}
Q_0&=&-|\Delta|^2-D^2\\
Q_1&=&\frac{iD\dot{\Delta}^*}{\Delta^*}-i\dot{D}\\
Q_2&=&-\frac{\ddot{\Delta}^*}{2\Delta^*}+\frac{3(\dot{\Delta}^*)^2}{4(\Delta^*)^2}.
\end{eqnarray}
Seeing the $\hbar$-dependence, one can see that the Stokes lines of the
above equation is described by 
\begin{eqnarray}
\label{eq-triv}
Q_0=|\Delta|^2+D^2,
\end{eqnarray}
which is exactly the same as the original Landau-Zener transition.
Therefore, for $D(t)=vt$ and $\Delta(t)=\Delta_0e^{-i\theta(t)}$, the
exponents are calculated by the same integral
($\int_{\gamma\pm}S_{odd}dt$), where the difference appears only in $S_{odd}$.
Here we calculate $S_{odd}$ explicitly for our later calculation,
\begin{eqnarray}
S_{odd,-1}&=&\sqrt{-D^2- |\Delta|^2}\\
S_{odd,0}&=&
\frac{i D\left(\frac{\dot{\Delta}^{*}}{\Delta^*}\right)-i \dot{D}}{2
\sqrt{-D^2-|\Delta|^2 }}\\
S_{odd,1}&=&\frac{F(x)}{32 \left(-|\Delta|^2-D^2\right)^{5/2}},
\end{eqnarray}
where the explicit form of $F(x)$ is
\begin{eqnarray}
F(x)&=&\frac{8 D^3 \dot{D} \dot{\Delta^{*}}}{\Delta^*}
+8 D \Delta \dot{D} \dot{\Delta^{*}}-20 D \Delta^* \dot{D} \dot{\Delta}\nonumber\\
&&-20 D \Delta  \dot{D} \dot{\Delta^*}+4 \Delta \Delta^* (\dot{D})^2
+8 D \Delta \Delta^* \ddot{D}\nonumber\\
&&-\frac{4 D^4 ((\Delta^*)')^2}{(\Delta^*)^2}
-\frac{4 D^2 \Delta  (\dot{\Delta^*})^2}{\Delta^*}
+8 D^2 \dot{\Delta} \dot{\Delta^*}\nonumber\\
&&+4 D^2 \Delta^* \ddot{\Delta}+\frac{12 D^4 (\dot{\Delta^*})^{2}}{(\Delta^*)^2}
+\frac{24 D^2 \Delta  (\dot{\Delta^*})^2}{\Delta^*}\nonumber\\
&&-\frac{8 D^4 \ddot{\Delta^*}}{\Delta^*}
-12 D^2 \Delta  \ddot{\Delta^{*}}-5 (\dot{\Delta^{*}})^2 (\dot{\Delta})^2\nonumber\\
&&-2 \Delta  \Delta^* \dot{\Delta} \dot{\Delta^{*}}+4 \Delta  (\Delta^*)^2 \ddot{\Delta}
+7 \Delta^2 (\dot{\Delta^{*}})^2\nonumber\\
&&-4 \Delta^2 \Delta^* \ddot{\Delta^*}
-16 D^2 (\dot{D})^2+8 D^3 \ddot{D}.
\end{eqnarray}
Although these calculations may seem tedious, they can all be carried
out with the aid of calculation software very easily.

We are now ready to perform all calculations.
However, as this paper mainly describes how to calculate the exponents
mechanically, it is recommended to check the original Berry's
papers\cite{Berry:1990, Berry:1984jv} to
understand the physical (geometric) meaning behind the calculations.
On the other hand, as mentioned in the introduction, the rigidity of
the approximation is not guaranteed in conventional calculations using
parallel transport, especially when non-perturbative effects come into play.
These ({\it both}) calculations are insufficient on their own,
and it is necessary for physics to consider both sides.

\subsection{The exact WKB for the twisted Landau-Zener transition}
The twisted Landau-Zener model is a model with the following Hamiltonian.
\begin{eqnarray}
\hat{H}(\tau)&=&{\bf H}(\tau)\cdot {\bf S}\nonumber\\
&\equiv &\left(
\begin{array}{cc}
Z(\tau) & X(\tau)-iY(\tau)\\
X(\tau)+iY(\tau)&-Z(\tau)
\end{array}
\right),
\end{eqnarray}
where $S$ is the vector spin$-\frac{1}{2}$ operator and ${\bf H}$ is
called the hamiltonian vector.
In particular, a model with
\begin{eqnarray}
{\bf H}(\tau)&=& \left(\Lambda \cos \phi(\tau), \Lambda \sin \phi(\tau),
		  A\tau\right)
\end{eqnarray}
and 
\begin{eqnarray}
\phi(\tau)&=&B\tau^2
\end{eqnarray}
appear to have a twisted trajectory because they seem to change
direction in the middle of the helical motion\cite{Berry:1990}.
In this model, the diagonal and off-diagonal elements are described by
\begin{eqnarray}
\Delta(\tau)&=&\Lambda e^{-i\phi(\tau)}\\
D(\tau)&=&A\tau,
\end{eqnarray}
where $\Delta(\tau)$ describes rotation on the $xy$-plane and $D(\tau)$
describes monotonic elevation in the z-direction.
The direction of the rotation changes at $\tau=0$, where the trajectory
seems to be ``twisted''.

First, we describe how to calculate exponents with the exact WKB for the 
basic model proposed by Berry\cite{Berry:1990}.
The dynamical exponent is calculated as
\begin{eqnarray}
\frac{1}{2}\int_{\gamma_\pm}S_{-1}d\tau&=& \pi i \left[
 \underset{\tau=\infty}{\Res}\sqrt{Q_0}\right]\nonumber\\
&=& \pi i \left[
 \underset{\tau=\infty}{\Res}\sqrt{-\Lambda^2-A^2\tau^2}\right]\nonumber\\
&=&\frac{\pi \Lambda^2}{2 A},
\end{eqnarray}
which (of course) coincides with Berry's calculation.
The geometric exponent for $A>0$ is
\begin{eqnarray}
\frac{1}{2}\int_{\gamma_\pm}S_{0}d\tau
&=&\pi i \underset{\tau=\infty}{\Res}
\left[
\frac{Q_1}{2\sqrt{Q_0}}
\right]\nonumber\\
&=& \pi i \underset{\tau=\infty}{\Res}
\left[
\frac{-A+2B\tau^2 i}{2\sqrt{A^2\tau^2+\Lambda^2}}
\right]\nonumber\\
&=&\frac{-B\Lambda^2\pi}{2A^2}+\frac{\pi}{2}i,
\end{eqnarray}
which is exact and coincides with the original
calculation\cite{Berry:1990}.

In addition to the dynamical exponent calculated above, there is a
higher term that does not trivially vanish in this model. 
We can calculate this term (quasi-geometric exponent) as
\begin{eqnarray}
\frac{1}{2}\int_{\gamma_\pm}S_{1}d\tau
&=&\pi i \underset{\tau=\infty}{\Res}
\left[
\frac{Q_2}{2\sqrt{Q_0}}-\frac{5}{32}\frac{Q_0^{'2}}{Q_0^{5/2}}
+\frac{Q_0^{''}-Q_1^2}{8Q_0^{3/2}}
\right]\nonumber\\
&=&\pi i \underset{\tau=\infty}{\Res}
\left[
\frac{
C_4 \tau^4 + C_3 \tau^3 +C_2 \tau^2 +
C_1 x +C_0}{32 \left(-A^2 \tau^2-\Lambda^2 \right)^{5/2}}
\right]\nonumber\\
&=&\frac{\pi  B^2 \Lambda^2}{2 A^3},
\end{eqnarray}
where
\begin{eqnarray}
C_4&=&16\left(- A^2 B^2 \Lambda^2+ i A^4 B - i A^4 B\right)\nonumber\\
C_3&=&8A^4\nonumber\\
C_2&=&-24 A^4 - 32 i A^2 B \Lambda^2 - 16 B^2 \Lambda^4 + 16 A^2 B \Lambda^2 i\nonumber\\
C_1&=&8A^2 \Lambda^2\nonumber\\
C_0&=&-4 A^2 \Lambda^2 - 16 i B \Lambda^4.
\end{eqnarray}
 As we have mentioned earlier, it can be easily proven that
any further exponent vanishes for 
this model.
The calculation with the exact WKB is very easy and powerful.

In this model, the twist and the intersection of the diagonal elements
are assumed to occur simultaneously at $\tau=0$, but one of them can be shifted
easily.
Now we have
\begin{eqnarray}
\Delta(\tau)&=&\Lambda e^{-i\phi(\tau)}\\
D(\tau)&=&A(\tau-\tau_0),
\end{eqnarray}
which gives the dynamical exponent,
\begin{eqnarray}
\frac{1}{2}\int_{\gamma_\pm}S_{-1}d\tau&=& \pi i \left[
 \underset{\tau=\infty}{\Res}\sqrt{Q_0}\right]\nonumber\\
&=& \pi i \left[
 \underset{\tau=\infty}{\Res}\sqrt{-\Lambda^2-A^2(\tau-\tau_0)^2}\right]\nonumber\\
&=&\frac{\pi \Lambda^2}{2 A},
\end{eqnarray}
which (of course) coincides with Berry's calculation.
The geometric exponent for $A>0$ is
\begin{eqnarray}
\frac{1}{2}\int_{\gamma_\pm}S_{0}d\tau
&=&\pi i \underset{\tau=\infty}{\Res}
\left[
\frac{Q_1}{2\sqrt{Q_0}}
\right]\nonumber\\
&=& \pi i \underset{\tau=\infty}{\Res}
\left[
\frac{-A+2B\tau^2 i}{2\sqrt{A^2(\tau^2-\tau_0)^2+\Lambda^2}}
\right]\nonumber\\
&=&\frac{-B\Lambda^2\pi}{2A^2}+\frac{\pi}{2}i,
\end{eqnarray}
which is exact and coincides with the original model.
It is immediately apparent that there is no change in the second factor
(quasi-geometric exponent) as well.
This shows (at least for this simple model) that {\it both} of the
geometric factors are 
not correlated with the dynamical process of the Landau-Zener transition.
Does this property still hold even if the model becomes more
complex?
It may be easy to predict it from the functions, but let us calculate the
exponents explicitly for a more complex model.

\subsection{Double twist model ($\phi(\tau)=\lambda \tau^3$)}
\label{sec-DT}
Here we consider a model in which the ``twist'' occurs twice
simultaneously and the direction of rotation is restored. 
If we assume $\phi(\tau)=\lambda \tau^3$, the dynamical exponent is
calculated as
\begin{eqnarray}
\frac{1}{2}\int_{\gamma_\pm}S_{-1}d\tau&=& \pi i \left[
 \underset{\tau=\infty}{\Res}\sqrt{Q_0}\right]
=\frac{\pi \Lambda^2}{2 A},
\end{eqnarray}
which (of course) coincides with the single twist($\phi(\tau)=A\tau^2$),
as $Q_0$ is identical to the single twist model.
On the other hand, for $A>0$ the geometric exponent is
\begin{eqnarray}
\frac{1}{2}\int_{\gamma_\pm}S_{0}d\tau
&=&\pi i \underset{\tau=\infty}{\Res}
\left[
\frac{Q_1}{2\sqrt{Q_0}}
\right]\nonumber\\
&=& \pi i \underset{\tau=\infty}{\Res}
\left[
\frac{i A \left(3 \lambda i \tau^3-1\right)}{2 \sqrt{-A^2 \tau^2-\Lambda^2}}
\right]\nonumber\\
&=&\frac{i\pi}{2},
\end{eqnarray}
which means $\Gamma_g=0$ in this case.
On the other hand, the quasi-geometric exponent is 
\begin{eqnarray}
\frac{1}{2}\int_{\gamma_\pm}S_{1}d\tau
&=&\pi i \underset{\tau=\infty}{\Res}
\left[
\frac{Q_2}{2\sqrt{Q_0}}-\frac{5}{32}\frac{Q_0^{'2}}{Q_0^{5/2}}
+\frac{Q_0^{''}-Q_1^2}{8Q_0^{3/2}}
\right]\nonumber\\
&=&-\frac{27 \pi  \lambda^2 \Lambda^4}{16 A^5},
\end{eqnarray}
which differs from the single twist and does not vanish in this model.
Again, it is easy to prove that higher exponents vanish for this model.

Next, we are going to check the correlation by shifting the position of
the twists and the intersection in this model.
For $D(\tau)=A(\tau-\tau_a)$, the dynamical exponent is calculated as
\begin{eqnarray}
\frac{1}{2}\int_{\gamma_\pm}S_{-1}d\tau&=& \pi i \left[
 \underset{\tau=\infty}{\Res}\sqrt{Q_0}\right]
=\frac{\pi \Lambda^2}{2 A},
\end{eqnarray}
which coincides with the above result.
Therefore the dynamical exponent is not correlated with the twists.
For $A>0$, the geometric exponent is
\begin{eqnarray}
\frac{1}{2}\int_{\gamma_\pm}S_{0}d\tau
&=&\pi i \underset{\tau=\infty}{\Res}
\left[
\frac{Q_1}{2\sqrt{Q_0}}
\right]\nonumber\\
&=& \pi i \underset{\tau=\infty}{\Res}
\left[
-\frac{i A \left(1-3 \lambda i \tau^2 (\tau-\tau_a)\right)}{2 \sqrt{-A^2 (\tau-\tau_a)^2-\Lambda^2}}
\right]\nonumber\\
&=&\frac{i\pi}{2}
-\frac{3 \pi  \tau_a \lambda \Lambda^2}{2 A^2},
\end{eqnarray}
which shows a clear correlation.
The geometric exponent disappears when the twists and the intersection
coincide at $\tau_a=0$.
The quasi-geometric exponent can also be calculated very easily, 
which becomes
\begin{eqnarray}
\frac{1}{2}\int_{\gamma_\pm}S_{1}d\tau
&=&\pi i \underset{\tau=\infty}{\Res}
\left[
\frac{Q_2}{2\sqrt{Q_0}}-\frac{5}{32}\frac{Q_0^{'2}}{Q_0^{5/2}}
+\frac{Q_0^{''}-Q_1^2}{8Q_0^{3/2}}
\right]\nonumber\\
&=&\frac{27 \pi  \lambda^2 \Lambda^2 \left(4 \tau_a^2 A^2 
		-  \Lambda^2 \right)}{16 A^5}.
\end{eqnarray}
These calculations show explicitly that it is the second-order
derivative of the angular function that is essential for the geometric
exponent, not the presence of the twist.
On the other hand, for the quasi-geometric exponent, the situation is
obviously different.
The cubic term comes into play even if its second derivative vanishes at
the Landau-Zener transition.

\subsection{Non-linear diagonal elements and phantom Landau-Zener
  transition}
\label{sec-NL}
\begin{figure}[ht]
\centering
\includegraphics[width=1.0\columnwidth]{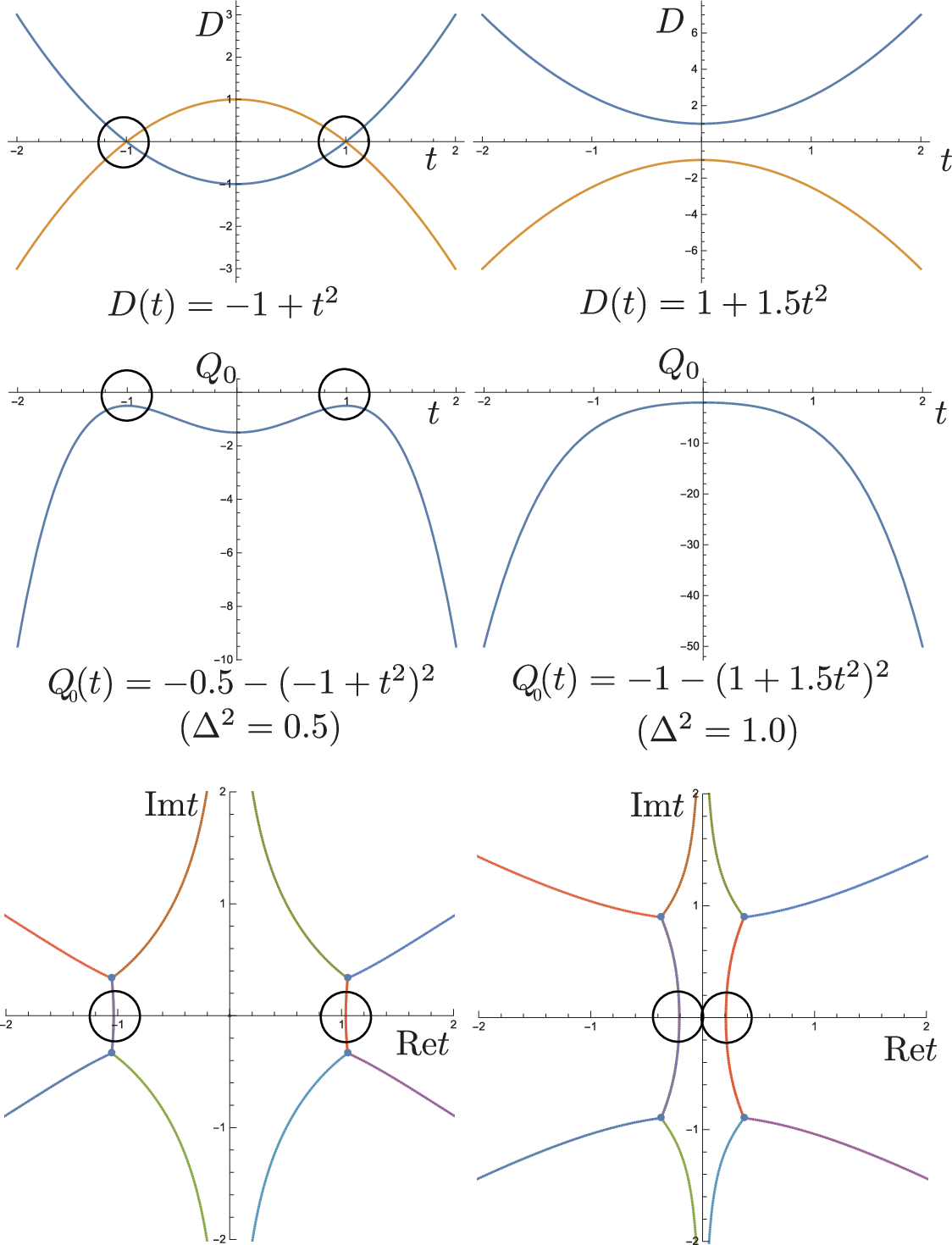}
 \caption{
The Stokes lines are written for non-linear diagonal elements
of the form $D(t)=\pm a^2+b^2 t^2$.
Two figures at the top show diagonal elements of the corresponding
 Landau-Zener model drawn on the $D-t$ plane, where $t$ is real.
The level crossings appear for $D(t)=- 1+ t^2$ (three graphs on the
 left), while they do not appear for $D(t)=1+1.5 t^2$ (right).
Two graphs in the middle show $Q_0(t)$ on the $Q-t$ plane for real $t$
 when the Landau-Zener model is converted to the standard form of the
 exact WKB; the graph on the left shows conventional ``hilltop''
 potential near the level crossings (marked by two circles), 
whereas the graph on the right does not show such a characteristic shape.
This means that the scattering point is not manifest on the right-hand
 case.
Two bottom figures show the Stokes lines of the model
 drawn on the complex $t$-plane;
the left graph shows conventional MTP structure near
 the level crossings, whereas the right graph shows the same
 MTP structure although there is no level crossing.
Given that the conventional Landau-Zener model is defined by level
 crossings, the situation depicted on the right could be termed
 the ``phantom Landau-Zener model''.}
\label{fig-quartic-stokes}
\end{figure}

\begin{figure}[ht]
\centering
\includegraphics[width=1.0\columnwidth]{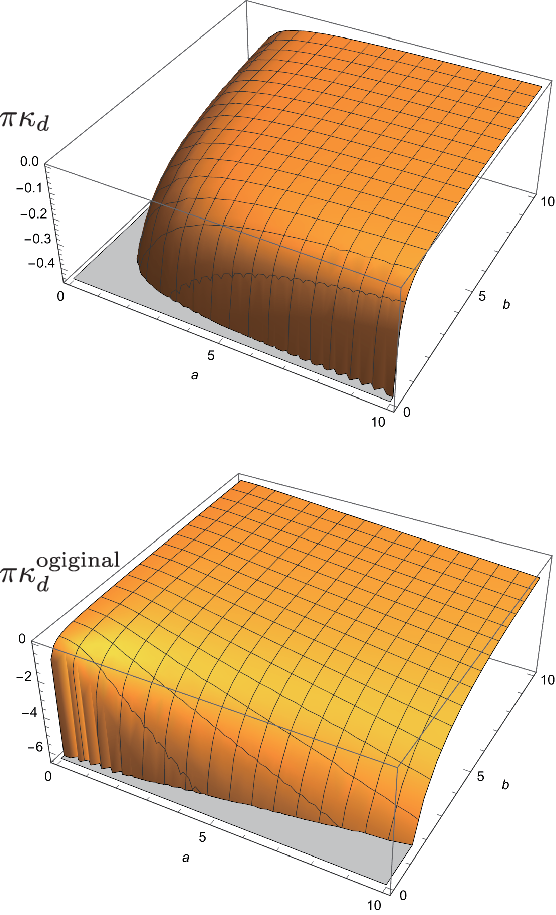}
 \caption{In the upper panel, the linear approximation given by
 Eq.(\ref{eq-kappa-linear}) is plotted on the $(a,b)$ plane assuming
 $\Delta=1$.
In the lower panel, the original (exact) equation (\ref{eq-kappa-ori}) is plotted. 
As is expected from the equations, one can see that the difference 
 in small $a$ is quite significant.}
\label{fig-kappa2}
\end{figure}

The simplest example will be analyzed here, for the case in which the
diagonal elements are not linear.
Specifically, we will discuss the computation of the dynamical 
exponent for $D(t)=\pm a^2+b^2 t^2$ and $\Delta=$const.
Looking at the situation shown in Fig.\ref{fig-quartic-stokes}, 
one will find that the situation is far from obvious.
Considering the behavior of $\pm D(t)$ in this case, 
they will either cross in two places or do not cross,
but two pairs of MTPs appear on the left and the right in both cases.
In the crossing case (upper row of Fig.\ref{fig-quartic-stokes}), if one
follows the argument of the original 
Landau-Zener transition\cite{Zener:1932ws}, one would consider the
Landau-Zener transitions twice as an approximation.
This result (with the conventional approximation) can be compared with
the result obtained by the exact WKB.
 See our later calculations and Fig.\ref{fig-kappa2}.
Contrary to our naive expectations, the results show that the situation
is far from obvious.

\begin{figure}[ht]
\centering
\includegraphics[width=1.0\columnwidth]{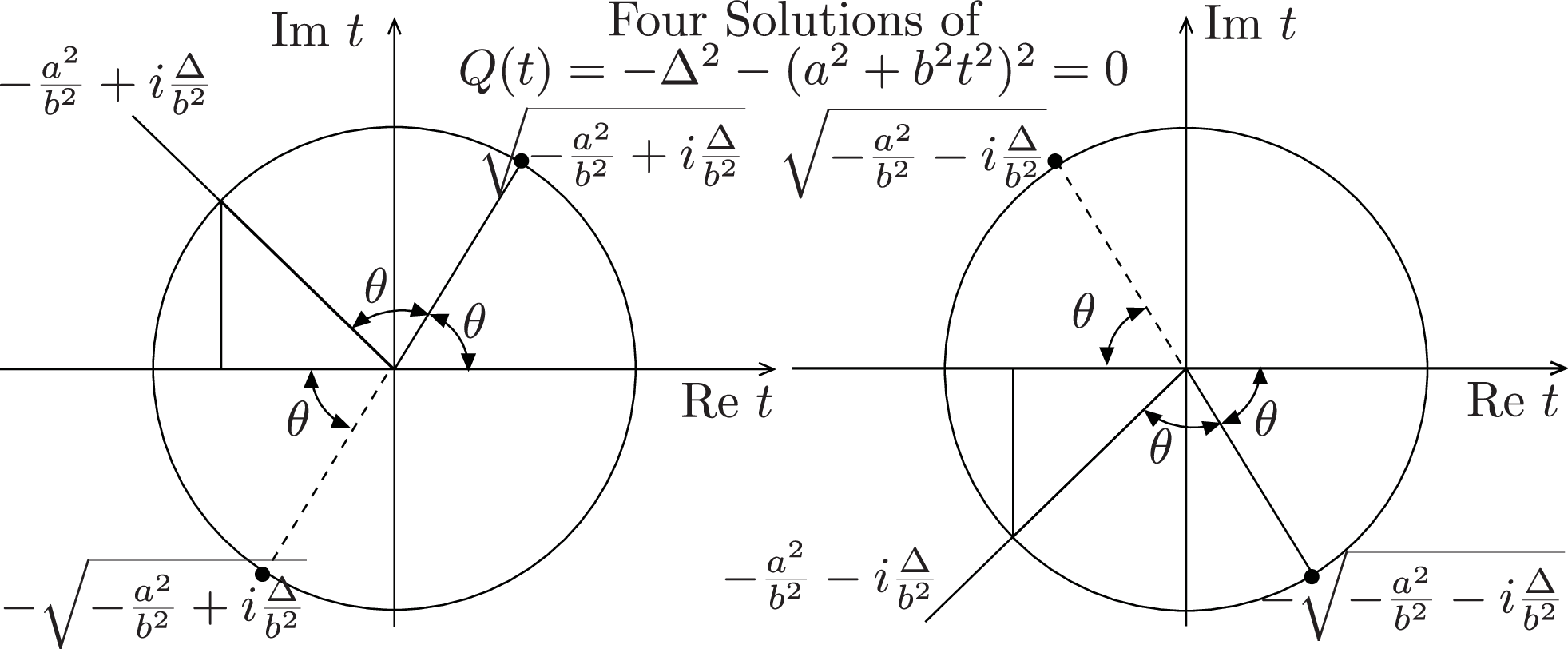}
 \caption{
Four solutions of $Q(t)=-\Delta^2-(a^2+b^2 t^2)^2$ 
on the complex $t$-plane appear symmetrically
 on either side. 
The two on the left-hand side and the two on the right-hand side are
 paired to form the Stokes lines of the MTP structure.
These are the solutions {\it without} crossing  $\pm D(t)$, but
the solutions with crossing $\pm D(t)$ are described in the same way.
Intuitively, the similarity of these solutions indicates that the Stokes
 lines will be similar.}  
\label{fig-4sol}
\end{figure}

Note that when there is no intersection, it is even more difficult to
find an appropriate approximation.
The original Landau-Zener model is analyzed at the intersection, so we
call the case without an intersection the ``phantom'' Landau-Zener transition.
If one writes the Stokes lines, one will notice that
almost the same structure of the Stokes phenomenon appears in both
cases.
How the turning points appear in the non-intersecting cases is illustrated in 
Fig.\ref{fig-4sol}.
By looking at the four turning points and the structure of the Stokes
lines, it can be seen that the Landau-Zener transition should be
considered in the same way near the left and right MTP structures,
regardless of whether $\pm D(t)$ intersects or not.
The problem is that in the non-intersecting case, the approximation
cannot be considered at the intersection.
This is not a situation that was considered in the original Landau-Zener
model.
Nevertheless, as the basic structure of the Stokes lines is the same, we can
show that the transfer matrix in each MTP structure will be
in the same form. 
As with the generalized Landau-Zener model, it is sufficient to
calculate the integral of $\int S_{odd}$, as the same MTP structures
appear (although they appear twice).

Let us now verify the validity of the conventional
linear approximation of the 
Landau-Zener transition by some concrete calculations.
We consider
\begin{eqnarray}
Q(t)&=&-\Delta^2-D(t)^2,\nonumber\\
&&D(t)=-a^2+b^2 t^2,
\end{eqnarray}
where $a,b,\Delta>0$ are real.
The intersections ($D(t)=0$) are found at
\begin{eqnarray}
t_{\pm}&=&\pm \frac{a}{b}.
\end{eqnarray}
The original Landau-Zener transition is considered at these
intersections using linear approximation.
On the other hand, the four turning points for the exact WKB are 
\begin{eqnarray}
t_{TP,R-}&=&-\frac{\sqrt{a^2-i\Delta}}{b}\\
t_{TP,L-}&=&-\frac{\sqrt{a^2+i\Delta}}{b}\\
t_{TP,R+}&=&\frac{\sqrt{a^2+i\Delta}}{b}\\
t_{TP,L+}&=&\frac{\sqrt{a^2-i\Delta}}{b},
\end{eqnarray}
where MTP pairs are $t_{TP,L\pm}$ and $t_{TP,R\pm}$.
To write the Landau-Zener transition at $t_+$ in the standard linear
form, we introduce $\tau\equiv t-a/b$ and write the transition at $t=t_+$
with $\tau=0$.
At $\tau=0$, we have the standard Landau-Zener transition given by
\begin{eqnarray}
D_+(\tau)&\equiv&D'(t_+)\tau=(2ab) \tau\\
Q_+(\tau)&\equiv&-\Delta^2-D_+(\tau)^2=-\Delta^2-4a^2b^2\tau^2.
\end{eqnarray}
The dynamical exponent of this transition is described as
\begin{eqnarray}
\label{eq-kappa-linear}
-\pi
 \kappa_d&=&\int_{-i\frac{\Delta}{2ab}}^{i\frac{\Delta}{2ab}}
\sqrt{-\Delta^2-4a^2b^2\tau^2} d\tau.
\end{eqnarray}
On the other hand, if we write the integral of the MTP structure using
the right pair of turning points, we find
\begin{eqnarray}
\label{eq-kappa-ori}
-\pi\kappa_d^{\mathrm{original}}&=&
\int_{t_{TP,R-}}^{t_{TP,R+}}
\sqrt{-\Delta^2-(a^2-b^2t^2)^2}dt,\nonumber\\
\end{eqnarray}
which may require additional conditions if one wants to justify the
conventional (linear) approximation at the intersection.
To examine the validity of the linear approximation, we plot 
$\kappa_d$ and $\kappa_d^{original}$ in Fig.\ref{fig-kappa2}.
As is obvious from the numerical calculation in Fig.\ref{fig-kappa2},
the conventional linear approximation is not reliable for this model,
especially for small $a$.
Using the exact WKB, the validity of the linear approximation can be
verified for more complex models.
Although the linear approximation at the intersection of the
Landau-Zener transition may appear to be quite conceivable, 
it turns out that it is not as easily valid as one might expect.
This would be particularly problematic if analytical calculations were
carried out using the linear approximation without numerical substitutions.

Furthermore, we have predicted from the Stokes lines that the
({\it phantom}) Landau-Zener transition does exist even in the {\it absence}
 of the intersection. 
In this case, since there is no suitable linear approximation, the
calculation of the exact WKB for the MTP structure would have to be used
in the first place.
The integrals can be calculated analytically following the protocol
described in this paper, but for non-linear models, they can no longer be
written down in terms of elementary
functions\cite{Voros:1983,Enomoto:2021hfv}.

It should be emphasized here that a major advantage of using the exact
WKB is that no matter how complex the model becomes, the 
systematic calculation can
be made by writing out the Stokes lines.\footnote{A simple Stokes-Line drawer
for Mathematica can be downloaded from Ref.\cite{Matsuda:download}.}
The Stokes line can also be used to find the phantom Landau-Zener
transition, which is not visible by the intersection.
The important point here is that changes in the global structure of
Stokes lines are closely linked to changes in the structure of the
theory.
For the same reason, there are cases where phenomena that appear to be
different are actually based on the same type of theory.
At first glance, the phantom Landau-Zener model seems different from the
Landau-Zener model with level crossings. However, an examination of the
Stokes line structure reveals that they are the same type.
As is explicitly shown in Fig.\ref{fig-kappa2}, the exact WKB calculation also revealed that linear approximation at the
level-crossing is far less accurate than commonly believed.

\subsection{Landau-Zener transition with $\Delta(\tau)=\Lambda
  e^{i\Theta(\tau)/\hbar}$}
\label{sec-Q}
Here, one naive question would arise for the exact WKB.
Since $\hbar$ is the parameter of the singular perturbation in
mathematics, the Planck constant should no longer be treated simply as a
``small parameter'' in the exact WKB. 
Is there a simple model in which we can see that the Planck constant
plays such a special role in the differential equation?

The Schwinger effect is known to be enhanced by oscillating electric
fields\cite{Taya:2020dco, Villalba-Chavez:2019jqp}. 
While ``oscillating electric field'' can be directly observed,
``photon wave function'' cannot. 
The significance of this difference on the
behavior of solutions to differential equations remains unclear.
The discussion presented here is not a solution to the problem, but it is
worth mentioning as an interesting phenomenon that sheds light on the
problem.

In two-level models such as the Landau-Zener transition, it is known
that the effect of the electromagnetic field can be different
when it behaves as a classical electromagnetic field or as a quantum 
photon\cite{TWO:2021, YXN:2008, XLFXY:2012}.
As we have briefly mentioned in our introduction, the effective
Hamiltonian for the 3D dirac fermion subject to rotating electric fields
can be casted to\cite{TWO:2021}
\begin{eqnarray}
\hat{H}&=&v \left[
\xi(k_x+e A_x)\hat{\sigma}^x
+(k_y+e A_y)\hat{\sigma}^y
+k_z \hat{\sigma}^z\right].\nonumber\\
\end{eqnarray}
Instead of using the electric field, one may use the magnetic field as
is described in Ref.\cite{NR:1994}.
Again, introducing $D(\tau)$ and $\Delta(\tau)$, we rephrase the model into a
simple form
\begin{eqnarray}
\label{eq-simpleoriginalLZ2}
i\hbar \frac{d}{d\tau}\left(
\begin{array}{c}
X\\
Y
\end{array}
\right)&=&\left(
\begin{array}{cc}
D(\tau) & \Delta(\tau)^*\\
\Delta(\tau) & -D(\tau)
\end{array}
\right)
\left(
\begin{array}{c}
X\\
Y
\end{array}
\right),
\end{eqnarray}
which describes the twisted and untwisted Landau-Zener models.

It would be meaningful to consider the following model,
assuming that the quantized electromagnetic field will have an explicit
Planck constant, 
\begin{eqnarray}
\Delta(\tau)&=&\Lambda e^{i\Theta(\tau)/\hbar},
\end{eqnarray}
where $\Theta(\tau)\equiv \epsilon \tau$ and $\epsilon (=\hbar \omega)$.
Here $\epsilon$ is assumed to be the fundamental quantity for the
quantized field. 
For a classical field, we will assume 
\begin{eqnarray}
\Delta(\tau)&=&\Lambda e^{i\omega \tau},
\end{eqnarray}
where the Planck constant does not appear.

When the field is quantized, we expect an amplification of the Landau-Zener
transition (i.e, the Schwinger effect) due to the photon absorption.
However, the effect of such a photon cannot be recovered  in the
conventional calculation.\footnote{See also the calculation in
Ref.\cite{NR:1994}, in which the problem is avoided by introducing a
unitary transformation. 
It is also possible to derive similar effects by adding modulation to
the oscillation\cite{TWO:2021}.
However, the serious problem of the previous calculations is that they
are based on vague assumptions as we have explained in detail in our
introduction.}
If the Planck constant is just a small number, it may seem ridiculous
that whether or not the Planck constant is written explicitly in the
differential equation does affect the physical quantity.
However, since the Planck constant is in fact not just a small
parameter of the equation but the key parameter of the singular
perturbation, it is not surprising that the 
nature of the solution varies depending on whether it appears in the
equation or not.

From the relation $\dot{\Delta}/\Delta=i\dot{\Theta}/\hbar$,
one will find for a ``quantized'' field as 
\begin{eqnarray}
Q_0&=&-[\Lambda^2+D^2]+D\Dot{\Theta}+\frac{1}{4}\Dot{\Theta}^2\\
Q_1&=&-iA-\frac{i}{2}\ddot{\Theta}\\
Q_2&=&0.
\end{eqnarray}
For $\Theta=\epsilon \tau$ and $D(\tau)=A\tau$, we find
\begin{eqnarray}
Q_0&=&-\left[\Lambda^2+A^2\tau^2-A\epsilon
	\tau-\frac{1}{4}\epsilon^2\right]\nonumber\\
&=&-\left[\Lambda^2+A^2\left(\tau-\frac{\epsilon}{2A}\right)^2-\frac{1}{2}\epsilon^2\right]\\
Q_1&=&-iA\\
Q_2&=&0.
\end{eqnarray}
Therefore, for the quantized field, the dynamical exponent is enhanced
by the term $\frac{1}{2}\epsilon^2$, which raises the
peak of the inverted quadratic potential of the scattering problem
described by the ``Schr\"odinger equation''\cite{TWO:2021}.
More simply, one can calculate the ``eigenvalues'' $E_\pm(\tau)$ to find that
the ``width'' $\Delta E(\tau)=|E_+(\tau)-E_-(\tau)|$ is all the same for
the 
conventional calculation, but for the above $Q_0$ one can see that there
is a significant change.
It may seem implausible at first glance that results can differ in the
exact WKB depending on whether the Planck constant is written 
explicitly or not in the equation.
Instead of thinking that the exact WKB is flawed, we should think that
the exact WKB provides an opportunity to understand how ``quantization''
appears in the fundamental equations.

\section{Conclusions and Discussions}
\label{sec-concdis}
\hspace*{\parindent}
In this paper, the geometric exponents are rigorously computed in models
in which non-perturbative effects are significant in non-adiabatic processes.
We calculated up to the next order of the conventional geometric
exponent (up to the quasi-geometric exponent) and showed that the exponents
will vanish in higher-order terms after the quasi-geometric exponent.
As shown in Ref.\cite{TWO:2021, Matsuda:2025hzn}, the Landau-Zener
transition is also used to describe the Schwinger effect, and various
applications of the Schwinger effect exist in condensed matter
physics\cite{TWO:2021, Berdyugin:2021njg, Taya:2020dco}.
Namely, in materials such as graphene, the ``vacuum'' corresponds to the
Dirac point, and applying a strong electric field induces electron-hole
pair creation by overcoming an effective energy
gap\cite{Berdyugin:2021njg}.
Also, the geometric factors have been actively studied in adiabatic
processes\cite{TKNN:1982, Uhlmann:1986}.
Our study provides a very effective method for mechanical and
rigorous calculations of the complex geometric factors in non-adiabatic
processes. 
We believe the convenience and
the accuracy of the calculation are greatly improved by the exact WKB.
We also analyzed the limits of the usual linear approximation and the 
``phantom'' Landau-Zener transition.
The phantom Landau-Zener transition is invisible as the
conventional linear approximation, but it has the same transition matrix
and the MTP structure.
Although this paper presents the exact WKB as a convenient computational 
tool, the exact WKB has a wide range of potential applications in
physics, as it refines complex phenomena such as
resonances\cite{Morikawa:2025xjq}, Non-perturbative Seiberg-Witten
geometry\cite{Alim:2022oll}, and non-Hermitian
systems\cite{Kamata:2023opn} and floquet system\cite{Fujimori:2025kkc}.
It also provides the local analyses for the Unruh effect and Hawking
radiation\cite{Matsuda:2025hzn}.
Non-perturbative effects are thought to become important when geometric
factors become discontinuous.
The discontinuity of the complex geometric factor is a universal phenomenon
that can appear in phase transitions, boundaries, particle
generation, and topology changes. 
We expect that the analysis of these non-perturbative effects using the
exact WKB will become increasingly important in the future.


\begin{thebibliography}{1}
\bibitem{WKB:review}
C.~M.~Bender and S.~A.~Orszag,
``Advanced Mathematical Methods for Scientists and Engineers'',
Springer (1978), ISBN-13:978-0387989310
\bibitem{Voros:1983}
A.~Voros, ``The return of the quartic oscillator -- The complex WKB
method'', Ann. Inst. Henri Poincare, 39 (1983), 211-338.
\bibitem{Delabaere:1993}
E. Delabaere, H. Dillinger and F. Pham: Resurgence de Voros et peeriodes
des courves hyperelliptique. Annales de l'Institut Fourier, 43 (1993), 163-
199.
\bibitem{Silverstone:2008}
H.~Shen and H.~J.~Silverstone, ``Observations on the JWKB treatment of
the quadratic barrier, Algebraic analysis of differential equations from
	microlocal analysis to exponential asymptotics'', Springer,
	2008, pp. 237 - 250.
\bibitem{Pham:1988}
F.~Pham, 
``Resurgence, quantized canonical transformations, and multiinstanton
expansions,''
Algebraic Analysis, Vol. II, Academie Press, 1988, pp. 699-726.
\bibitem{CNP:1993}
B.~Candelpergher, J.~G.~Nositicls and F.~Pham,
``Approche de la Resurgence,''
Hermann, 1993.
\bibitem{DDP:1993}
E.~Delabaere, H.~Dillinger and F.~Pham, 
``Resurgence de Voros et periodes des courbes hyper elliptiques,''
 Ann. Inst. Fourier (Grenoble) 4 3 (1993), 163-199.
\bibitem{DDP:1997}
E.~Delabaere, H.~Dillinger and F.~Pham,
``Exact semi-classical expansions for one dimensional quantum oscillators,''
J. Math. Phys. 38 (1997), 6126-6184.
\bibitem{ExactWKB:textbook}
T.~Kawai and Y.~Takei,
``Algebraic Analysis of Singular Perturbation Theory,''
Iwanami Series in Modern Mathematics, 2005, 978-0-8218-3547-0.
\bibitem{Virtual:2015HKT}
N.~Honda, T.~Kawai and Y.~Takei,
``Virtual Turning Points'', Springer (2015),  978-4-431-55702-9.
\bibitem{RPN:2017}
``Resurgence, Physics and Numbers'' edited by F.~Fauvet, D.~Manchon,
	S.~Marmi and  D.~Sauzin, Publications of the Scuola Normale
	Superiore, ISBN-13:978-88-7642-613-1.
\bibitem{Bouchard:2017}
V. ~Bouchard and B. ~Eynard, ``Reconstructing WKB from topological
recursion'', Journal de l' Ecole polytechnique - Mathematiques,
Volume 4 (2017), 845-908.
\bibitem{Enomoto:2020xlf}
S.~Enomoto and T.~Matsuda,
``The exact WKB for cosmological particle production,''
JHEP \textbf{03} (2021), 090.
\bibitem{Enomoto:2021hfv}
S.~Enomoto and T.~Matsuda,
``The exact WKB and the Landau-Zener transition for asymmetry in cosmological particle production,''
JHEP \textbf{02} (2022), 131.
\bibitem{Enomoto:2022nuj}
S.~Enomoto and T.~Matsuda,
``The Exact WKB analysis for asymmetric scalar preheating,''
JHEP \textbf{01} (2023), 088
\bibitem{Berry:1984jv}
M.~V.~Berry,
``Quantal phase factors accompanying adiabatic changes,''
Proc. Roy. Soc. Lond. A \textbf{392} (1984), 45-57
\bibitem{Berry:1990}
M.~V.~Berry, 
``Geometric amplitude factors in adiabatic quantum transitions'',
Proc. R. Soc. Lond. A \textbf{430}(1990),405-411
\bibitem{XCN:2010}
D.~Xiao, M-C~Chang, and Q.~Niu
``Berry phase effects on electronic properties'',
Rev. Mod. Phys. \textbf{82} (2010) 1959 
\bibitem{TKNN:1982}
D.~J.~Thouless, M.~Kohmoto, M.~P.~Nightingale, and M.~den~Nijs,
``Quantized Hall Conductance in a Two-Dimensional Periodic Potential'',
Phys. Rev. Lett. \textbf{49}, 405
\bibitem{Uhlmann:1986}
A.~Uhlmann,
``Parallel transport and ``quantum holonomy'' along density operators'',
Rep. Math. Phys. \textbf{24}(1986) 229.
\bibitem{Geometric:textbooks}
Q.~.Niu, Ming-Che~Chang, B.~Wu, D.~Xiao, R.~Cheng,
``Physical Effects of Geometric Phases'',
 World Scientific (2017), ISBN-13:978-981-3225-72-5
\bibitem{Aharonov:1987gg}
Y.~Aharonov and J.~Anandan,
``Phase Change During a Cyclic Quantum Evolution,''
Phys. Rev. Lett. \textbf{58} (1987), 1593
\bibitem{Wilczek:1984dh}
F.~Wilczek and A.~Zee,
``Appearance of Gauge Structure in Simple Dynamical Systems,''
Phys. Rev. Lett. \textbf{52} (1984), 2111-2114
\bibitem{Samuel:1988zz}
J.~Samuel and R.~Bhandari,
``General Setting for Berry's Phase,''
Phys. Rev. Lett. \textbf{60} (1988), 2339-2342
\bibitem{Mukunda:1991rc}
N.~Mukunda and R.~Simon,
``Quantum kinematic approach to the geometric phase. 1. General formalism,''
Annals Phys. \textbf{228} (1993), 205-268
\bibitem{Bohm:2003}
A.~Bohm, A.~Mostafazadeh, H.~Koizumi, Q.~Niu, J.~Zwanziger,
``The Geometric Phase in Quantum Systems
Foundations, Mathematical Concepts, and Applications in Molecular and
	Condensed Matter Physics'',
Springer (2003), ISBN-13:978-3-642-05504-1
\bibitem{Matsuda:download}
A simple Stokes-Line drawer for mathematica,\\
https://www.sit.ac.jp/user/matsuda/img/EWKB1.zip.
\bibitem{NR:1994}
K.~Nakamura and S.~A.~Rice,
``Nonadiabatic transitions and gauge structure'',
Phys. Rev. A \textbf{49} R2217(1994)
\bibitem{BKDSW:1996}
D.~Bouwmeestert, G.~P.~Karman, N.~H.~Dekker, C.~A.~Schrama, and
	J.~P.~Woerdman,
``Observation of the Geometric Amplitude
	Factor in an Optical System'',
Journal of Modern Optics \textbf{43}(1996)10,2087-2103
\bibitem{TWO:2021}
S.~Takayoshi, J.~Wu and T.~Oka,
``Nonadiabatic nonlinear optics and quantum geometry ? Application to
	the twisted Schwinger effect'',
SciPost Phys. \textbf{11}  (2021) 075
\bibitem{YXN:2008}
W.~Yao, D.~Xiao and Q.~Nin, 
``Valley-dependent optoelectronics from inversion symmetry breaking,
	Phys. Rev. B \textbf{77} (2008) 235406
\bibitem{XLFXY:2012}
D.~Xiao, G.-B.~Liu, W.~Feng, X.~Xu and W.~Yao,
``Coupled spin and valley physics in monolayers of $MoS_2$ and other
group-vi dechalcogenides'',
Phys. Rev. Lett. \textbf{108} (2012) 196802 
\bibitem{Chung:2024lky}
D.~J.~H.~Chung and N.~Sudhir,
``Topological contribution to the Bogoliubov coefficient for cosmological particle production,''
Phys. Rev. D \textbf{111} (2025) no.10, 105009
\bibitem{Zener:1932ws}
  C.~Zener,
  ``Nonadiabatic crossing of energy levels,''
  Proc.\ Roy.\ Soc.\ Lond.\ A {\bf 137} (1932) 696.
\bibitem{Landau:1932vnv}
L.~D.~Landau,
``A theory of energy transfer. 2.,''
Phys. Z. Sowjetunion \textbf{2} (1932), 46
\bibitem{Dykhne:1962}
A.~M.~Dykhne,
``Adiabatic Perturbation of Discrete Spectrum States'',
JETP \textbf{14}(1962) 941.
\bibitem{Davis:1976}
J.~P.~Davis and P.~Pechukas,
``Nonadiabatic transitions induced by a time-dpendent Hamiltonian in the
	semiclassical/adiabatic limit: The two-state case''
J. Chem. Phys. \textbf{64} (1976) 3129
\bibitem{Matsuda:2025hzn}
T.~Matsuda,
``Quantum field theory on curved manifolds,''
[arXiv:2501.09919 [hep-th]].
\bibitem{Berdyugin:2021njg}
A.~I.~Berdyugin, N.~Xin, H.~Gao, S.~Slizovskiy, Z.~Dong,
	S.~Bhattacharjee, P.~Kumaravadivel, S.~Xu, L.~A.~Ponomarenko and
	M.~Holwill, \textit{et al.}
``Out-of-equilibrium criticalities in graphene superlattices,''
Science \textbf{375} (2022) no.6579, 430-433
\bibitem{Taya:2020dco}
H.~Taya, T.~Fujimori, T.~Misumi, M.~Nitta and N.~Sakai,
``Exact WKB analysis of the vacuum pair production by time-dependent
	electric fields,''
JHEP \textbf{03} (2021), 082
\bibitem{Morikawa:2025xjq}
O.~Morikawa and S.~Ogawa,
``Unified exact WKB framework for resonance --
	Zel'dovich/complex-scaling regularization and rigged Hilbert
	space,'' 
[arXiv:2505.02301 [hep-th]].
\bibitem{Alim:2022oll}
M.~Alim, L.~Hollands and I.~Tulli,
``Quantum Curves, Resurgence and Exact WKB,''
SIGMA \textbf{19} (2023), 009
\bibitem{Kamata:2023opn}
S.~Kamata,
``Exact WKB analysis for PT-symmetric quantum mechanics: Study of the Ai-Bender-Sarkar conjecture,''
Phys. Rev. D \textbf{109} (2024) no.8, 085023
\bibitem{Fujimori:2025kkc}
T.~Fujimori, S.~Kamata, T.~Misumi, N.~Sueishi and H.~Taya,
``Exact-WKB Analysis of Two-level Floquet Systems,''
\bibitem{Voros-ASU:2022}
T.~Aoki, T.~Suzuki and S.~Uchida,
``An elementary proof of the Voros connection formula for WKB solutions
	to the Airy equation with a large parameter'', arXiv:2205.02988
\bibitem{Voros-AU:2022}
T.~Aoki and S.~Uchida,
``Voros Coefficients at the Origin and at the Infinity of the
	Generalized Hypergeometric Differential Equations with a Large
	Parameter'', arXiv:2104.13751
\bibitem{Villalba-Chavez:2019jqp}
S.~Villalba-Ch{\'a}vez and C.~M{\"u}ller,
``Signatures of the Schwinger mechanism assisted by a fast-oscillating electric field,''
Phys. Rev. D \textbf{100} (2019) no.11, 116018
\end{thebibliography}
\end{document}